\documentclass[showpacs,amsmath,amssymb,aps,showkeys,floatfix,prc,onecolumn,a4paper,preprint]{revtex4-2}
\usepackage{graphicx}
\usepackage{epstopdf}
\usepackage{dcolumn}
\usepackage{bm}
\usepackage{epsfig}
\usepackage{amsfonts}
\usepackage{amssymb,amscd}
\usepackage{hyperref}
\usepackage{xcolor}
\hypersetup{
    colorlinks=true,                
    breaklinks=true,                
    urlcolor= black,                
    linkcolor= blue,                
    bookmarksopen=false,
    filecolor=black,
    citecolor=blue,
    linkbordercolor=blue,
    pdfborder = {0 1 0}
}

\begin{document}

\title{Charmed meson production based on dipole transverse momentum representation in high energy hadron-hadron collisions available at the LHC}

\pacs{12.38.-t; 13.60.Le; 13.60.Hb}

\author{G. Sampaio dos Santos}
\affiliation{High Energy Physics Phenomenology Group, GFPAE  IF-UFRGS \\
Caixa Postal 15051, CEP 91501-970, Porto Alegre, RS, Brazil}

\author{G. Gil da Silveira}
\affiliation{CERN, PH Department, 1211 Geneva, Switzerland}
\affiliation{High Energy Physics Phenomenology Group, GFPAE  IF-UFRGS \\
Caixa Postal 15051, CEP 91501-970, Porto Alegre, RS, Brazil}
\affiliation{Departamento de F\'{\i}sica Nuclear e de Altas Energias, Universidade do Estado do Rio de Janeiro\\
CEP 20550-013, Rio de Janeiro, RJ, Brazil}

\author{M. V. T. Machado}
\affiliation{High Energy Physics Phenomenology Group, GFPAE  IF-UFRGS \\
Caixa Postal 15051, CEP 91501-970, Porto Alegre, RS, Brazil}

\begin{abstract}
The production of $D$-mesons in high-energy $pp$ collisions at the LHC kinematic regime is analyzed with the dipole approach in the momentum representation. We present predictions for the $D$-meson differential cross section in terms of the transverse momentum and rapidity distributions taking into account the nonlinear behavior of the QCD dynamics. Comparison between our results and the corresponding experimental measurements reported by the ALICE and LHCb Collaborations in different rapidity bins is performed. We show that the $D$-meson production in the high energy limit can be properly addressed by using the QCD dipole transverse momentum distributions.
\end{abstract}

\maketitle

\section{Introduction}
\label{intro}

The study of charmed mesons has been performed over the years, resulting in improvements on both experimental and theoretical fields. For instance, the production mechanism and the properties of such mesons are subjects of investigation. The charmed mesons are the lightest particles that have a heavy quark as its constituent, consequently, they can be a relevant tool to test theoretical frameworks regarding quarks and their interactions \cite{godwiss}. In particular, $D$-mesons were first observed in 1976 by experimental measurements performed by the SLAC-LBL Collaboration with the Mark I detector at the SPEAR collider at center-of-mass energies from 3.9 to 4.6~GeV \cite{goldhaber,peruzzi}. 
The $D$-meson production was investigated in $e^{+}e^{-}$ annihilation as well as in deep inelastic $ep$ scattering (DIS), where the process directly probes the gluon distribution in the proton. Aiming the scenario of the heavy ion programme at high energy colliders, the study of the $D$-meson in $pp$ collisions serves as a powerful baseline to investigate the cold nuclear matter effects as well as the effects originated in hot matter by a medium known as quark-gluon plasma (QGP) \cite{andronic,prino}. However, considering hadron-hadron collisions in the high energy regime, charm quarks are produced in the hard scattering processes between the initial-state partons present in the colliding hadrons. Subsequently, there is the hadronization process of these heavy quarks originating $D$-mesons in the final state. In particular, with the advent of the Large Hadron Collider (LHC) and further developments -- specially concerning the precision of the measurements plus a wide windows of center-of-mass energy, transverse momentum, and rapidity -- provide an interesting scenario to study charmed meson production. The transverse momentum and rapidity distributions probed at the LHC energies allow to investigate the $D$-meson production at small values of the Bjorken variable $x$, where significant nonlinear effects of the QCD regimes are expected. Therefore, the processes involving the open meson production is expected to be sensitive to the nonlinear QCD dynamics. Hence, this particular kinematic region can be accessed with $D$-meson measurements at forward rapidities.
The production cross section dependent on the detector kinematic variables is obtained by the scope of QCD calculations. The standard approaches consist of obtaining the function of the squared momentum transfer $Q^2$, known as collinear factorization \cite{collins}, or in terms of the partonic transverse momentum $k_T$, namely $k_T$-factorization \cite{gribov, march,levin,catani} formalism. For example, studies concerning heavy quarks in conjunction with heavy $D$ mesons assuming collinear factorization can be found in the literature based on the general-mass variable-flavour-number scheme (GM-VFNS) \cite{kniehl,helenius}, and in the fixed order plus next-to-leading logarithms approach (FONLL) \cite{cacciari,cacciari1}. On the other hand, calculations for heavy quark production in the $k_{T}$-factorization framework are available in Refs.~\cite{ryskin,kharz,shab,lusz,mac,chachamis}. Additionally, an analysis regarding the $D$-meson production including the intrinsic heavy quark component in the hadron wave function is performed in Refs.~\cite{vicnav,carv}.

The $k_T$-factorization approach is applied to processes in hadron-hadron scattering at the high-energy limit and, in such framework, hard scattering matrix elements at small-$x$ are calculated. In particular, one employs the gluon densities  via the unintegrated gluon distribution (UGD). The gluon primordial
transverse momentum distribution  allows evaluating and extracting information concerning the properties of the structure of the proton as well as the QCD evolution equations that take into account the transverse momentum of the partons. Moreover, the UGD is also a function of the momentum fraction $x$ and the factorization scale $\mu_{F}^2$. Such densities are not computed from first principles and it then need to be parametrized. There are in the literature various models for UGD which differ on underlying assumptions. Then, those observables strongly sensitive to the UGD need to be investigated in order to constrain those $k_T$-dependent distributions \cite{angeles}. Within the $k_T$-factorization framework, the $D$-meson production cross section is obtained by the corresponding charm quark production process described in terms of UGD at small-$x$ at the scale $\mu_F\sim 2m_c$.

The $D$-meson production in high-energy processes (equivalently small-$x$) can be investigated within the color dipole formalism \cite{nik}, which has been proven suitable to evaluate different processes in high energy phenomenology. In the color dipole framework, the phenomenology is based on the universal dipole cross section that includes the nonlinear behavior and high order corrections of the QCD dynamics \cite{rauf}. In such framework the hard process is viewed in terms of $q\bar{q}$ dipole scattering off the target. Namely, the projectile emits a gluon that subsequently fluctuates into a $q\bar{q}$ color pair characterizing  a color dipole with definite size which interacts with the color field of the target. The corresponding dipole amplitude is related to the intrinsic dipole $k_T$-distribution, i.e the dipole transverse momentum distribution (TMD). In the limit of large gluon transverse momentum the dipole TMD corresponds approximately to the UGD. In the present investigation we consider analytical expressions for the TMDs based on parton saturation physics. 

In this work, based on the theoretical scenario of the dipole approach in transverse momentum representation, we perform predictions for the $D$-meson production focusing on high energy $pp$ collisions at the LHC. Moreover, our results take into account large and low $p_T$-spectrum by considering a wide range of rapidity bins. Both forward and central rapidities are considered for $D^0$, $D^+$, and $D^{*+}$ production as well as the cross section ratios $\sigma (D^+)/\sigma (D^0)$ and $\sigma (D^{*+})/\sigma (D^0)$ at central rapidities. The main novelty is the use of the recently proposed  phenomenological parameterization for the UGD based on geometric scaling properties that correctly reproduces the hadron spectrum in $pp$ collisions \cite{mpm}. It describes the saturated and dilute perturbative QCD regimes and has been successfully extended to heavy ion collisions \cite{Moriggi:2020qla,Peccini:2021rbt}. Moreover, we consider a simplified  “Weizsäcker–Williams” (WW) gluon TMD which has been used in studies of $Z^0$ hadroproduction \cite{ww}.

The paper is organized as follows. In Sec.~\ref{form} the theoretical framework to obtain the $D$-meson production in the  dipole formalism in transverse momentum representation  is presented. In Sec.~\ref{res} results are shown for several analytical models for the gluon TMD that are compared to the experimental measurements reported by the ALICE and LHCb Collaborations at the LHC, with the corresponding theoretical uncertainties investigated.  At last, in Sec.~\ref{conc} we summarize our main conclusions and remarks. 

\section{Theoretical formalism}
\label{form}

The charmed meson production is evaluated within the QCD dipole framework, where the basic assumption consists that the production process can be determined by a color dipole, $Q\bar{Q}$, interacting with the nucleon/nucleus in the target rest frame. The inclusive production of a $Q\bar{Q}$ -- originated from virtual gluon fluctuation -- is written in terms of the cross section of the process $g+N\rightarrow Q\bar{Q}+X$, including the superposition of color-singlet and color-octet contributions as well. The hadronic   cross section of the process $pp \rightarrow QX$ assumes the form
\begin{eqnarray} 
\frac{d^{4}\sigma_{pp \rightarrow QX}}{dy d\alpha d^2p_T} = F_{g}(x_1,\mu_{F}^2)\,
\frac{d^{3}\sigma_{gp \rightarrow QX}}{d\alpha d^2p_T} \,, 
\label{hdeq}
\end{eqnarray}
where $y$ and $p_T$ correspond to the rapidity and transverse momentum of the heavy quark (denoted as $Q$), respectively, and $\alpha$ 
$(\bar{\alpha} = 1 - \alpha)$ is the gluon momentum fraction exchanged with the heavy quark (antiquark). In the expression above the $gp \rightarrow Q\bar{Q}X$ cross section has been convoluted with the projectile gluon UGD. Ignoring the primordial gluon momentum, the quantity $F_{g}(x_1,\mu_{F}^2)$ is given by
\begin{eqnarray}
F_{g}(x_1,\mu_{F}^2) = \int^{\mu^2_F}\frac{dk_T^2}{k_T^2}{\cal{F}}(x_1,k_T^2)\,.
\label{fg}
\end{eqnarray}

The cross section, as computed in Eq.~(\ref{hdeq}), takes similar form used in the 
$k_T$-factorization framework.  The heavy quark TMD can be obtained in the momentum representation in terms of the dipole TMD, ${\cal T}_{\mathrm{dip}}$ \cite{vic}, in the following way:
\begin{eqnarray} 
\frac{d^3\sigma_{gp \rightarrow QX}}{d\alpha d^2 p_T } &=& 
\frac{1}{6\pi} \int \frac{d^2 \kappa_{\perp}}{\kappa^{4}_{\perp}}  \alpha_s(\mu_{F}^2)\, {\cal T}_{\mathrm{dip}}(x_2,\kappa^{2}_{\perp})\,\bigg\{\bigg[\frac{9}{8}{\cal{I}}_0(\alpha,\bar{\alpha},p_T) - \frac{9}{4} {\cal{I}}_1(\alpha,\bar{\alpha},\vec{p}_T,\vec{\kappa}_{\perp}) \nonumber \\ 
&+& {\cal{I}}_2(\alpha,\bar{\alpha},\vec{p}_T,\vec{\kappa}_{\perp}) + \frac{1}{8}{\cal{I}}_3(\alpha,\bar{\alpha},\vec{p}_T,\vec{\kappa}_{\perp})\bigg] + \left[\alpha \longleftrightarrow \bar{\alpha}\right]\bigg\} \,,  
\label{proxs} 
\end{eqnarray}
where $\alpha_s(\mu_{F}^2)$ stands for the running coupling in the one-loop approximation. Also, we have that the auxiliary quantities ${\cal{I}}_i$ ($i=0,1,2,3$) are given by:
\begin{eqnarray} 
{\cal{I}}_0(\alpha,\bar{\alpha},p_T) &=& \frac{m_{Q}^2 + (\alpha^2 + \bar{\alpha}^2)p_{T}^2}{(p_{T}^2 + m_{Q}^2)^2} \label{I1} \,,\\
{\cal{I}}_1(\alpha,\bar{\alpha},\vec{p}_T,\vec{\kappa}_{\perp}) &=& \frac{m_{Q}^2 + (\alpha^2 + \bar{\alpha}^2) \vec{p}_T\cdot 
(\vec{p}_T - \alpha \vec{\kappa}_{\perp})}{[(\vec{p}_T - \alpha \vec{\kappa}_{\perp})^2 + m_{Q}^2](p_{T}^2 + m_{Q}^2)} \label{I2}
\,,  \\
{\cal{I}}_2(\alpha,\bar{\alpha},\vec{p}_T,\vec{\kappa}_{\perp}) &=& \frac{m_{Q}^2 + (\alpha^2 + \bar{\alpha}^2) 
(\vec{p}_T - \alpha \vec{\kappa}_{\perp})^2}{[(\vec{p}_T - \alpha \vec{\kappa}_{\perp})^2 + m_{Q}^2]^2} \label{I3}
\,,  \\
{\cal{I}}_3(\alpha,\bar{\alpha},\vec{p}_T,\vec{\kappa}_{\perp}) &=& \frac{m_{Q}^2 + (\alpha^2 + \bar{\alpha}^2) 
(\vec{p}_T + \alpha \vec{\kappa}_{\perp})\cdot (\vec{p}_T - \bar{\alpha} \vec{\kappa}_{\perp})}
{[(\vec{p}_T + \alpha \vec{\kappa}_{\perp})^2 + m_{Q}^2][(\vec{p}_T - \bar{\alpha} \vec{\kappa}_{\perp})^2 + m_{Q}^2]} \label{I4}\,,
\end{eqnarray}
with $m_Q$ being the heavy quark mass. Moreover, the projectile and target fractional light-cone momentum are denoted by  $x_{1}$ and $x_{2}$, respectively.  They are explicitly written in terms of the pair rapidity, $x_{1,2} = \frac{M_{Q\bar{Q}}}{\sqrt{s}}\, e^{\pm y}$, where $\sqrt{s}$ is the collision center-of-mass energy and $M_{Q\bar{Q}}$ represents the invariant mass of the $Q\bar{Q}$ pair,
\begin{eqnarray}
M_{Q\bar{Q}}\simeq 2\sqrt{m_{Q}^2+p_T^2}\,.
\label{invmass}
\end{eqnarray}
Furthermore, in Eq.~(\ref{proxs}) ${\cal T}_{\mathrm{dip}}(x_2,\kappa_{\perp}^2)$ is the intrinsic dipole TMD that is connected with the dipole cross section $\sigma_{q\bar{q}}$ by means of \cite{nik1, bart}
\begin{eqnarray} 
\sigma_{q\bar{q}}(\vec{r},x) \equiv \frac{2\pi}{3}\int \frac{d^2\kappa_{\perp}}{\kappa_{\perp}^4}\,
\big(1-e^{i\vec{\kappa}_{\perp} \cdot \, \vec{r}}\big)\big(1-e^{-i\vec{\kappa}_{\perp} \cdot \, \vec{r}}\big)\,{\cal T}_{\mathrm{dip}}(x,\kappa_{\perp}^2) \,.
\label{dip_UGD}
\end{eqnarray}
When the transverse momentum of the gluon target is large enough, such that $\kappa_\perp \gg \Lambda_{\rm QCD}$, a relation between the $k_\perp$-factorization and the dipole approach can be established implying that the intrinsic dipole TMD can be written approximately in terms of the UGD function times $\alpha_s$. In the $D$-meson production we can safely apply this approximated relation since that the heavy quark pair production is coupled with small-sized dipoles; this is validated by the range of heavy quark transverse momentum probed experimentally, consequently, 
\begin{eqnarray} 
{\cal T}_{\mathrm{dip}}(x_2,\kappa_{\perp}^2) \simeq \alpha_s\,{\cal F}(x_2,\kappa_{\perp}^2)\,,
\label{dip_kt}
\end{eqnarray}
where ${\cal F}(x_2,k_{T}^2)$ accounts for the target UGD. It is important to stress that the relation (\ref{dip_kt}) is not necessarily in the small $\kappa_\perp$ region, which is associated to dipoles of large sizes and where the gluon UGD is not sufficiently constrained.

As pointed previously, there exist several parametrizations for the UGD and here we will use the analytical models proposed in Refs.~\cite{gbw,mpm,ww}. The first model for gluon UGD  is derived from a saturated form of the Golec-Biernat and W\"usthoff (GBW) dipole cross section\cite{gbw} that effective accounts for a scattering of a color dipole off a nucleon \cite{gbw1}
\begin{eqnarray}
\sigma_{q\bar{q}}(r,x) = \sigma_0\, \left[1 - \mathrm{exp}\left(-\frac{r^2\,Q_s^2}{4}\right)\right]\,,
\label{dipGBW}
\end{eqnarray}
and, by applying the corresponding Fourier transform of the Eq.~(\ref{dip_UGD}), one arrives the expression:
\begin{eqnarray}
{\cal F}_{GBW}(x,k_{T}^2)=\frac{3\,\sigma_{0}}{4 \pi^2\alpha_{s}} \frac{k_{T}^4}{Q_{s}^2}\,\mathrm{exp}\left(-\frac{k_{T}^2}{Q_{s}^2}\right) \,,
\label{FGBW}
\end{eqnarray}
where $\alpha_{s} = 0.2$ and $Q_s$ is the saturation scale, $Q_{s}^2(x) = (x_0/x)^{\lambda}\, \mathrm{GeV^2}$. In this study we use the set of parameters $\sigma_{0}$, $x_{0}$, and $\lambda$ that has been determined from the fit done to the recently extracted data on $F_2$ at low-$x$ given in Ref.~\cite{gbwfit}.
Using the GBW UGD, the quantity $F_{g}(x_1,\mu_{F}^2) $ in Eq.~(\ref{fg}) can be analytically computed and reads
\begin{eqnarray}
F_{g}^{GBW}(x_1,\mu_{F}^2)  = \frac{3\sigma_0}{4\pi^2\alpha_s}Q_s^2(x_1)\left[ 1-\left(1+\frac{\mu_{F}^2}{Q_s^2(x_1)} \right)\exp\left(-\frac{\mu_{F}^2}{Q_s^2(x_1)} \right) \right]\,.
\end{eqnarray}

The second model has been recently proposed by Moriggi, Peccini, and Machado (MPM) \cite{mpm} taking into account the geometric scaling observed in high $p_T$ hadron production in $pp$ collisions along with a Tsallis-like behavior of measured hadron spectrum, given by:
\begin{eqnarray}
{\cal F}_{MPM}(x, k_T^2)=\frac{3\,\sigma_{0}}{4\pi^2\alpha_{s}}\frac{(1+\delta n)}{Q_{s}^2}
\frac{k_{T}^4}{\left(1+\frac{k_{T}^2}{Q_{s}^2} \right)^{(2+\delta n)}}\,,
\label{FMPM}
\end{eqnarray}
with the scaling variable being $\tau = k_{T}^2/Q_{s}^2$ and $Q_{s}^2(x) = (x_0/x)^{0.33}$. The function $\delta n = a\tau^{b}$ defines the powerlike behavior of the spectrum of the produced gluons at high momentum. The parameters $\sigma_{0}$, $x_{0}$, $a$, and $b$ are obtained by fitting DIS data at small-$x$. Moreover, the same value of the strong coupling used in GBW model is considered here.

Finally, the last model is the WW model \cite{ww} for the gluon distribution, inspired by the Weizs\"acker-Williams method of virtual quanta, considering the one-gluon exchange between a pointlike parton and a hard probe at large momentum transfer. This gluon exchange plays a role similar with the virtual photon exchange such that the associated virtual gluon density resembles the virtual photon density originated from a pointlike charge described by Weizs\"acker-Williams approximation. The UGD in this parametrization read as
\begin{eqnarray}
{\cal F}_{WW}(x,k_T^2) = 
\begin{cases}
{(N_1k_{T}^2/k_0 ^2)} {(1-x)^7 \, (x^\lambda k_T^2 / k_0 ^2)^{-b}} \quad \mbox{for } \; k_T^2 \geq k_0 ^2, \\
{(N_1k_{T}^2/k_0 ^2)} {(1-x)^7 \, x^{-\lambda b}}  \quad \mbox{for } \; k_T^2 < k_0 ^2,
\end{cases}
\label{FWW}
\end{eqnarray}
where the normalization constant $N_1 = 0.6$, $k_0 = 1$~GeV, and $\lambda = 0.29$. The factor $(1 - x)^7$ accounts for the gluon distribution suppression at large $x$ while the phenomenological parameter $b$ is responsible for controlling the $k_T$ scaling of the gluon distribution. Parametrization above has been used in studies of Lam–Tung relation breaking at the $Z^0$ hadroproduction in the context of $k_T$-factorization formalism. It was shown that the shape of WW TMD is crucial for the right description of that relation breaking.

With the purpose of obtaining the $D$-meson spectra, one has necessarily to assume the hadronization of the heavy quarks via the corresponding fragmentation function, which is interpreted as the probability that a heavy quark fragments into a given heavy meson. Therefore, the $D$-meson production can be calculated by convoluting the heavy quark cross section with the fragmentation function,
\begin{eqnarray} 
\frac{d^{3}\sigma_{pp \rightarrow DX}}{dY d^2P_T} = \int_{z_{\mathrm{min}}}^1 \frac{dz}{z^2} 
D_{Q/D} (z,\mu_{F}^2) \int_{\alpha_{\mathrm{min}}}^1 d\alpha \frac{d^{4}\sigma_{pp \rightarrow QX}}{dyd\alpha d^2p_T} \,,
\label{Dmes}
\end{eqnarray}
where $z$ is the fractional momentum of the heavy quark $Q$ carried by the $D$-meson and $D_{Q/D}(z,\mu_{F}^2)$ is the fragmentation function. Here  we will assume the parametrization proposed in Ref.~\cite{kkks08} that includes the DGLAP evolution. Moreover, the quantities $m_D$, $Y=y$, and $P_T$ are the mass, rapidity, and transverse momentum of the $D$-meson, respectively \cite{maciula}. As $z \equiv p_{T}/P_{T}$, one can use $p_{T} = P_{T}/z$ and the lower limits for the $z$ and $\alpha$ integrations are given by:
\begin{eqnarray}
z_{\mathrm{min}}&=&\frac{\sqrt{m_{D}^2+P_T^2}}{\sqrt{s}}\,e^{Y}\label{zmin} \,, \\ 
\alpha_{\mathrm{min}}&=&\frac{z_{\mathrm{min}}}{z}\sqrt{\frac{m_{Q}^2 z^2 + P_{T}^2}{m_{D}^2 + P_{T}^2}} \label{almin} \,.
\end{eqnarray}

The GBW parametrization allows us to obtain an approximate expression for the rapidity and $p_T$ distributions. In the kinematic range considered here the hard scale $\mu_F$ is higher than the saturation scale, $\mu_F^2/Q_s(x)\gg 1$. In this limit, $F_g^{GBW}\approx 3\sigma_0Q_s(x_1)/(2\pi)^2\alpha_s$. Moreover, at central rapidity, $Y=0$, the typical value of $z_{min}$ in the range $p_T<3m_D$ at $\sqrt{s}=5$~TeV is $z_{min}\sim 2\times 10^{-3}$. Based on this fact, the lower limit in the $\alpha$-integration can be safely considered $\alpha_{min}\rightarrow 0$. It can be shown that the heavy quark $p_T$-spectrum is given by:
\begin{eqnarray}
\frac{d^2\sigma (gp\rightarrow QX)}{d^2p_T}\approx \frac{3}{5}\frac{\sigma_0 Q_s^ 2(x_2)}{4(2\pi)^2}\left[\frac{p_T^4+\frac{25}{9}m_c^2p_T^2 +m_c^4}{(m_c^2+p_T^2)^4} \right]\,.
\end{eqnarray}

Instead of integrating over $z$ in Eq.~(\ref{Dmes}), we will compute the cross section using a simplification for the fragmentation function, $D_c(z,\mu_F)\sim \delta (z-\langle z\rangle_c)$. The average momentum fraction $\langle z\rangle_c $ is defined as \cite{kkks08},
\begin{eqnarray}
 \langle z\rangle_c (\mu_F) = \frac{1}{B_c(\mu_F)}\int_{z_{cut}}^{1}dz z D_c(z,\mu_F), \quad \mathrm{with} \quad B_c(\mu_F) = \int_{z_{cut}}^{1}dz D_c(z,\mu_F), 
\end{eqnarray}
where $B_c$ is the branching fraction $c \to D$ and $x_{cut}=0.1$ \cite{kkks08}. For the KKKS fragmentation function considered here, one has $\langle z\rangle_c (\mu_F=2m_c) = 0.573, 0.571$, and $0.617$ for $D^0$, $D^+$ and $D^{*+}$, respectively. The average fraction is weakly dependent on the hard scale $\mu_F$, with a $\sim 20\%$ decreasing at $\mu_F=m_Z$. Therefore, we will take $\langle z\rangle \equiv \langle z\rangle_c (2m_c)$ and the meson spectrum will be given by
\begin{eqnarray}
\frac{d^{3}\sigma_{pp \rightarrow DX}}{dY d^2P_T} \approx \left[\frac{\langle z\rangle \,\sigma_0 }{2(2\pi)^2}\right]^2\frac{Q_s^2(x_1)Q_s^2(x_2)}{5}\left[ \frac{9m_c^4\langle z\rangle^4 +25m_c^2\langle z\rangle^2P_T^2 + 9P_T^4}{(m_c^2\langle z\rangle^2+P_T^2)^4}\right].
\label{approximation}
\end{eqnarray}

In what follows we take the previously discussed UGD parametrizations to calculate the $D$-meson production in $pp$ collisions and performed a comparison with the respective experimental measurements obtained at the LHC.

\section{Results and discussions}
\label{res}

\begin{figure*}[!t]
\begin{tabular}{cc}
\includegraphics[scale=0.35]{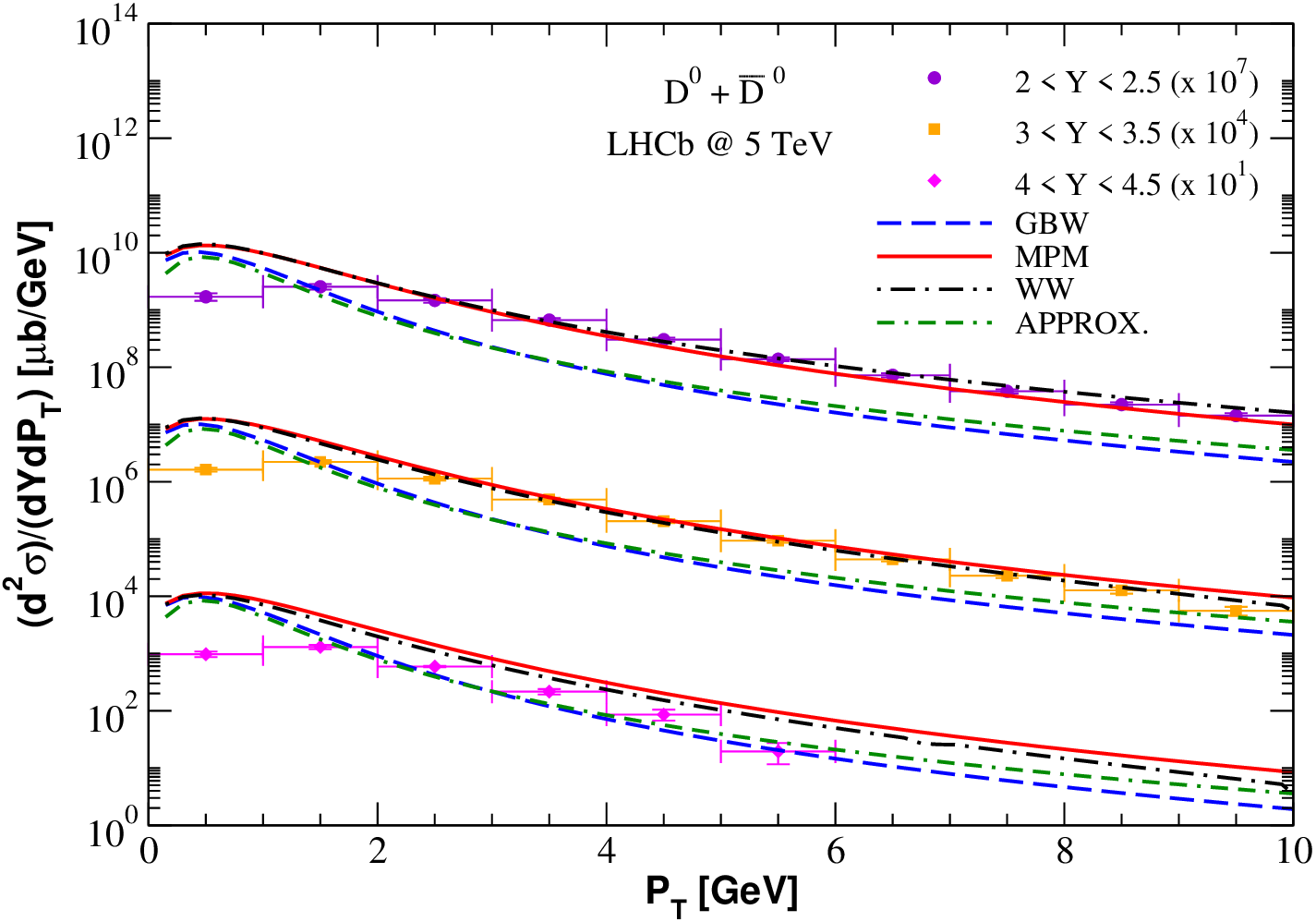}  \includegraphics[scale=0.35]{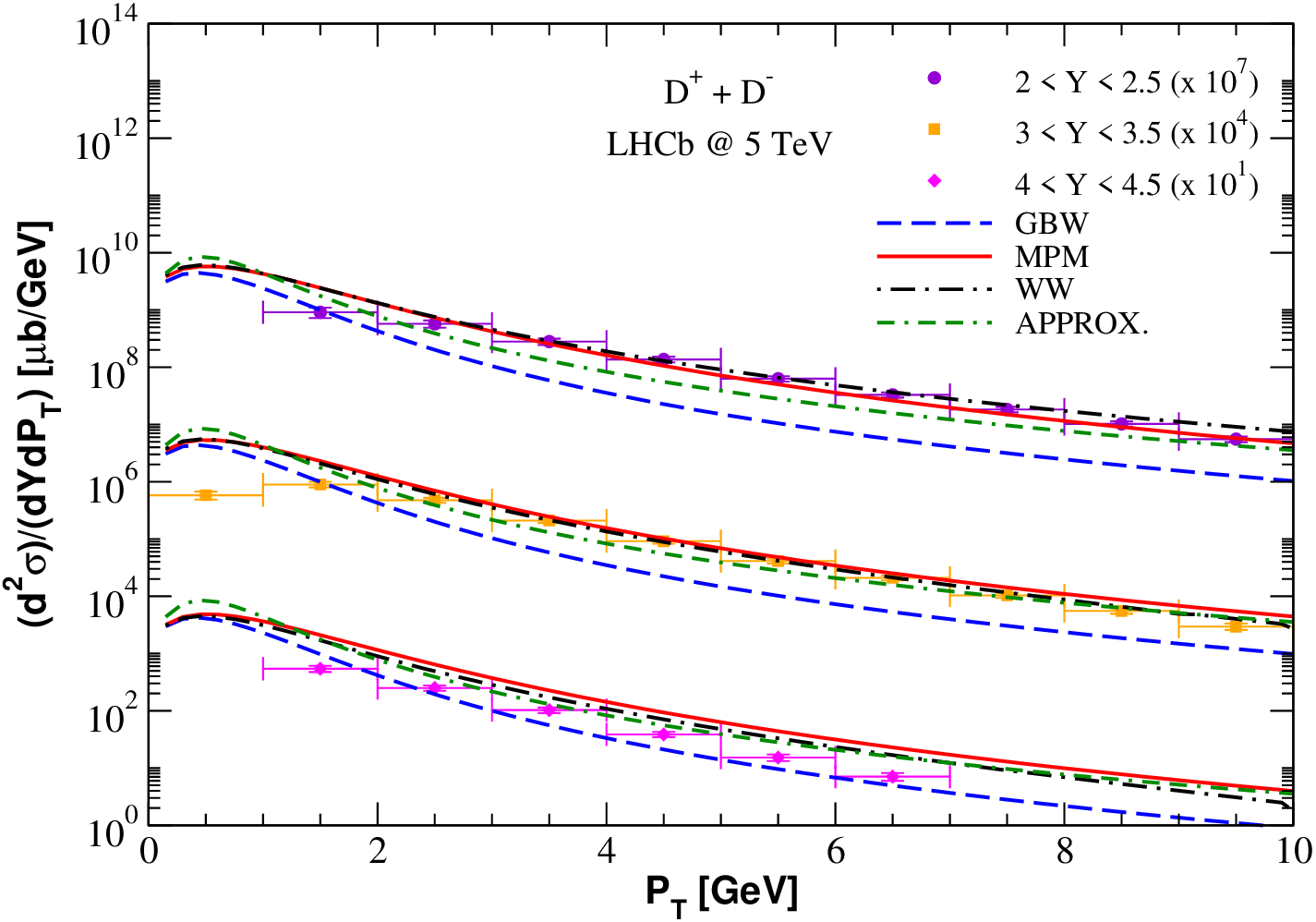} \\
\includegraphics[scale=0.35]{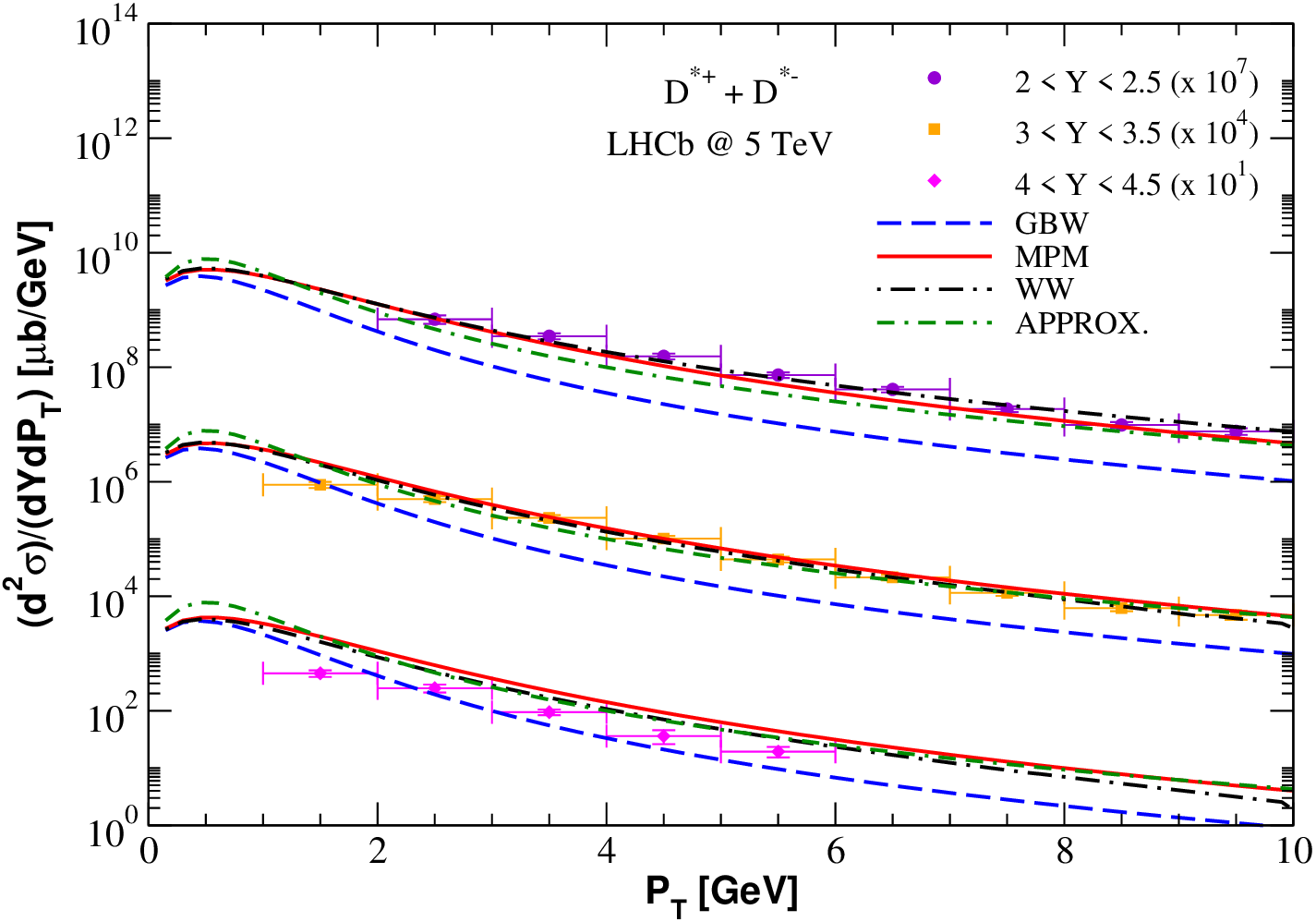}  
\end{tabular}
\caption{Double-differential cross sections for $D^0$ (left panel), $D^+$ (right panel) and $D^{*+}$ (bottom panel) production in $pp$ collisions at $\sqrt{s} = 5$~TeV considering three forward rapidity bins. The results are obtained using the GBW, MPM, and WW parametrizations as well as the approximate relation obtained in Eq.~\ref{approximation}. The corresponding comparison is performed with the measurements from the LHCb experiment \cite{LHCb5}.}
\label{pp5}
\end{figure*}

Let us present the results obtained with the dipole approach in transverse momentum representation  and using three parametrizations for the UGD introduced before, namely the GBW, MPM, and WW models. Considering the $D$-meson production in high energy hadron-hadron collisions we predict the distributions in transverse momentum and rapidity focusing in the LHC kinematic regime. Our results are directly compared to the experimental data reported by ALICE and LHCb  Collaborations.

In Fig.~\ref{pp5} we show the results for $D^0$, $D^+$, and $D^{*+}$ production including the charge conjugates states in $pp$ collisions at $\sqrt{s} = 5$~TeV. The predictions for the differential cross section are confronted against the measurements from the LHCb Collaboration \cite{LHCb5} considering three distinct rapidity bins: $2 < Y < 2.5$, $3 < Y < 3.5$, and $4 < Y < 4.5$. Selecting all the three $D$-meson and rapidity bins considered here, we can verify that the MPM and WW parametrizations give quite similar results at $P_T < 4$~GeV and both models are in good agreement with the experimental measurements, except for $P_T < 2$~GeV. Moreover, a slightly difference between MPM and WW results appears taking the spectrum from $P_T > 4$~GeV. This difference is a slightly more pronounced at very forward rapidity interval, $4 < Y < 4.5$, where such models are not in completely agreement to the correct normalization of the $P_T$ spectrum. In contrast, the GBW parameterization describe the experimental measurements in a narrow $P_T$ distribution, $2 < P_T < 3$~GeV, where it provides a better agreement at very forward rapidity. Apart from the particular $P_T$ spectrum mentioned before, the GBW results is loosing adherence to data. The reason for this behavior consists in the Gaussian shape present in the GBW approach that enters in Eqs.~(\ref{dip_kt}) and (\ref{FGBW}), leading to the suppression pattern observed in the results. For sake of comparison, we show the results taking into account the approximate expression for the $D$-meson spectrum given in Eq.~\ref{approximation} (labeled APPROX hereafter). For $D^0$ case, only the measurement at $4 < Y < 4.5$ can be reasonably described. However, considering the $D^{+}$ and $D^{*+}$ production, the predictions are in good agreement with the experimental data at the region $3 < Y < 3.5$. In these case, the approximate expression mimics the results from MPM or the WW UGDs. 

\begin{figure*}[!t]
\begin{tabular}{cc}
\includegraphics[scale=0.35]{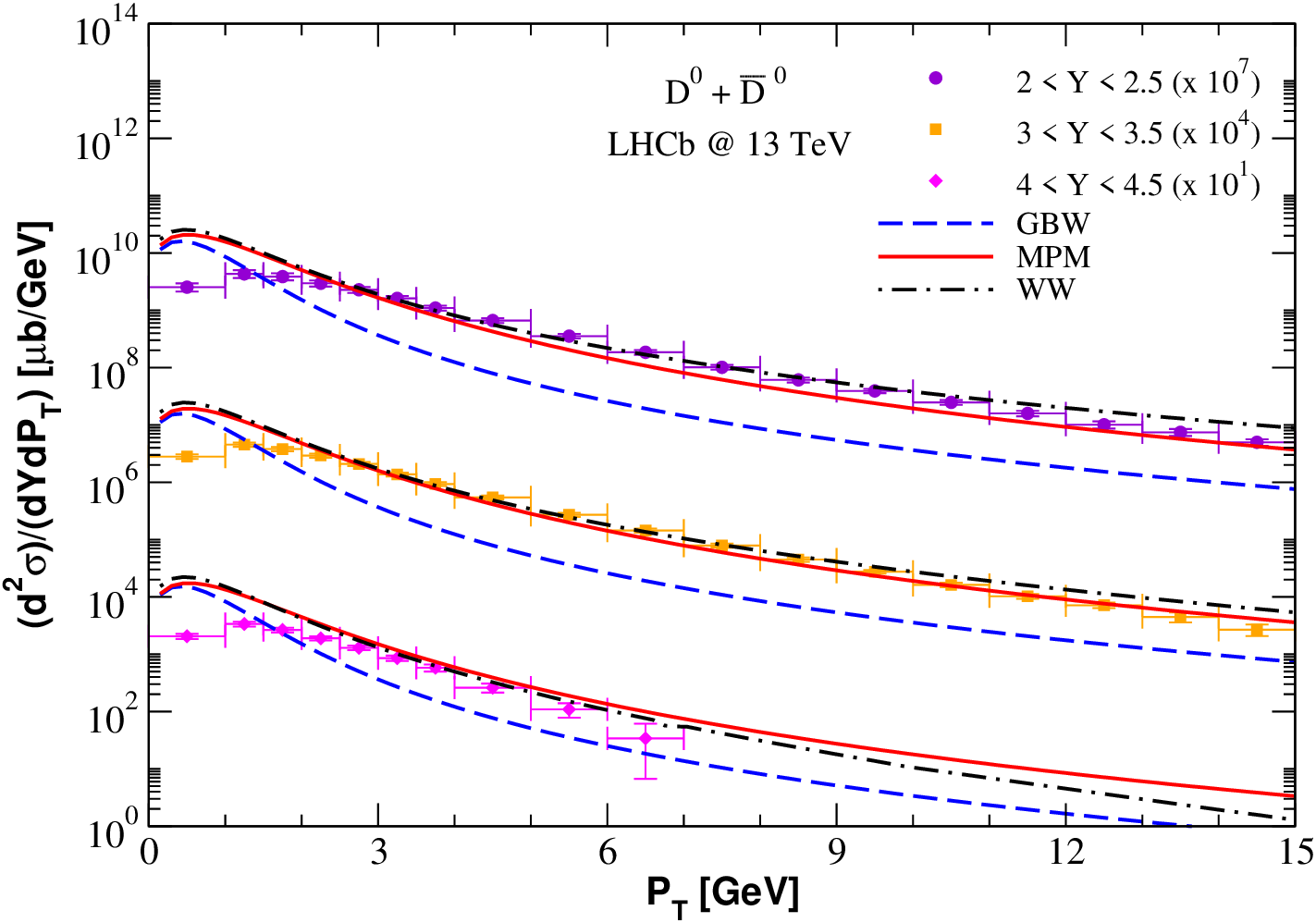}  \includegraphics[scale=0.35]{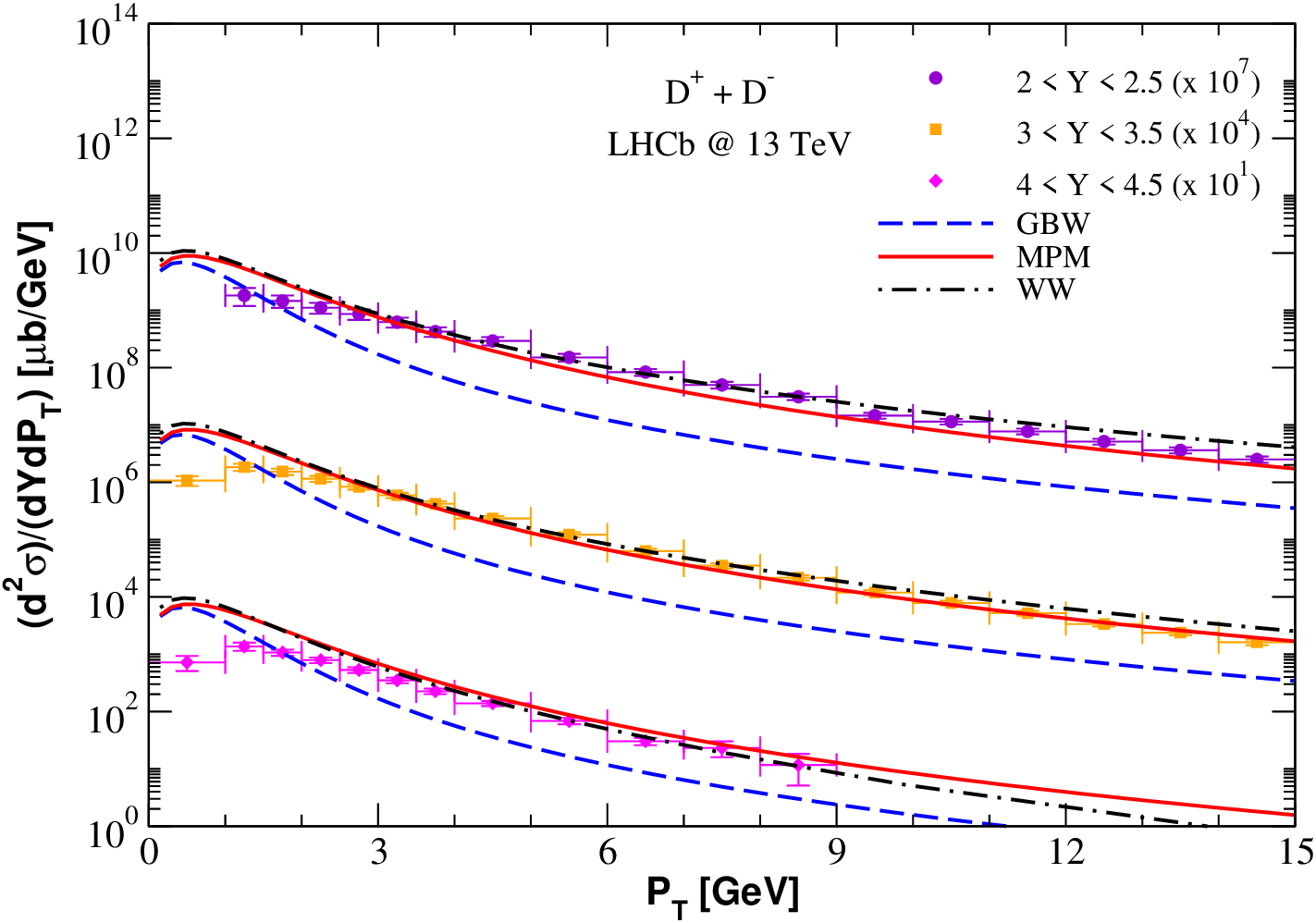} \\
\includegraphics[scale=0.35]{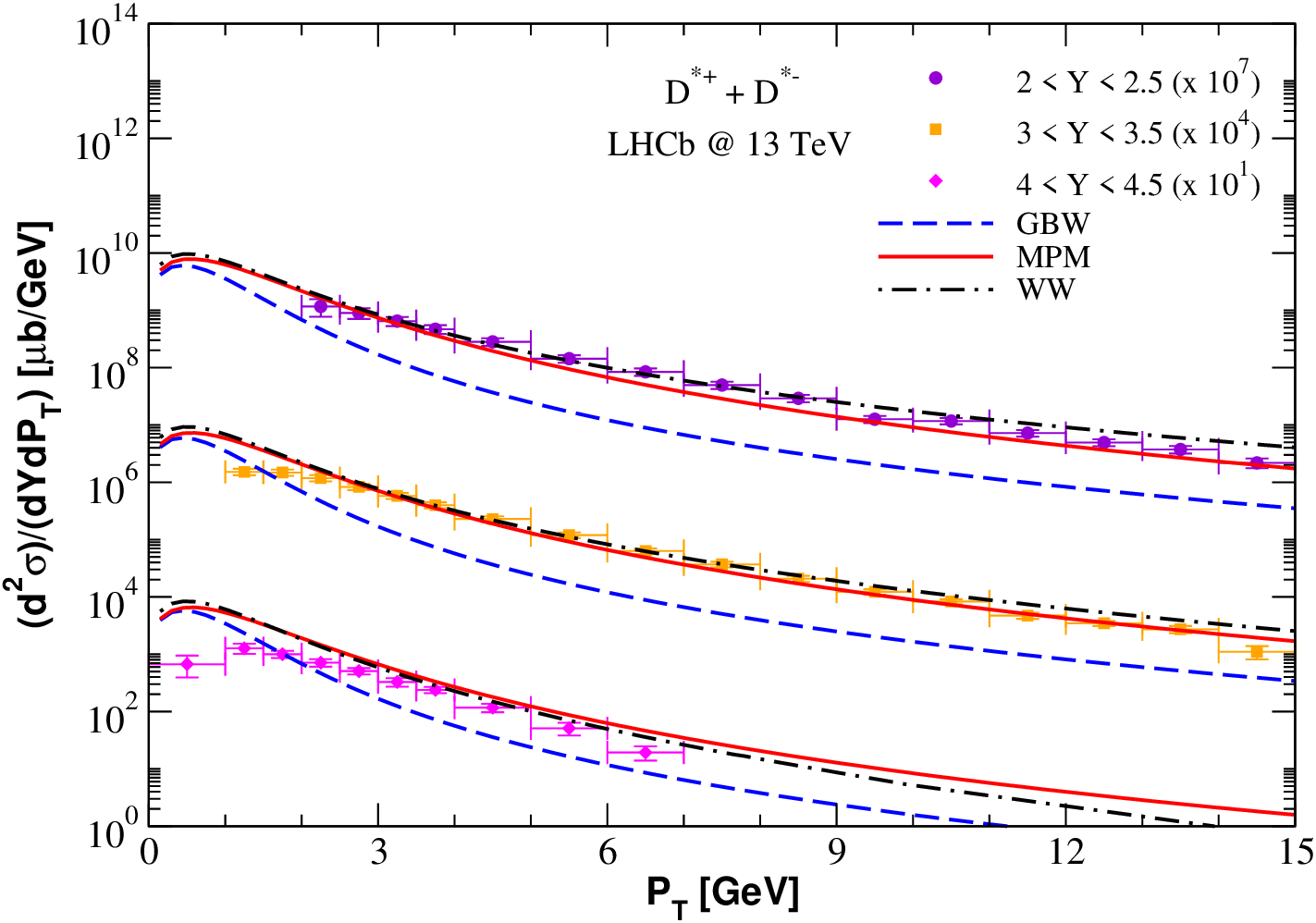}  
\end{tabular}
\caption{Double-differential cross sections for $D^0$ (left panel), $D^+$ (right panel), and $D^{*+}$ (bottom panel) production in $pp$ collisions at $\sqrt{s} = 13$~TeV considering three forward rapidity bins. The results are obtained using the GBW, MPM, and WW parametrizations and compared to the experimental measurements from the LHCb experiment \cite{LHCb13}.}
\label{pp13}
\end{figure*}

In Fig.~\ref{pp13} we present the numerical results that consider the same $D$-mesons and rapidity bins analyzed previously, but now at higher center-of-mass energy of $\sqrt{s} = 13$ TeV, with experimental data provided by the LHCb experiment \cite{LHCb13}. We can notice the same pattern observed  at $\sqrt{s} = 5$~TeV. However, the predictions with MPM and WW present some difference as the $P_T$ spectrum increases, specially considering the $2 < Y < 2.5$ and $4 < Y < 4.5$ rapidity bins. Moreover, the MPM and WW parametrizations present a significant improvement concerning the description of the experimental data in the very forward rapidity kinematic region. Additionally, we found that the same similarities with GBW results at $\sqrt{s} = 5$~TeV.

\begin{figure*}[!t]
\begin{tabular}{c}
\includegraphics[scale=0.35]{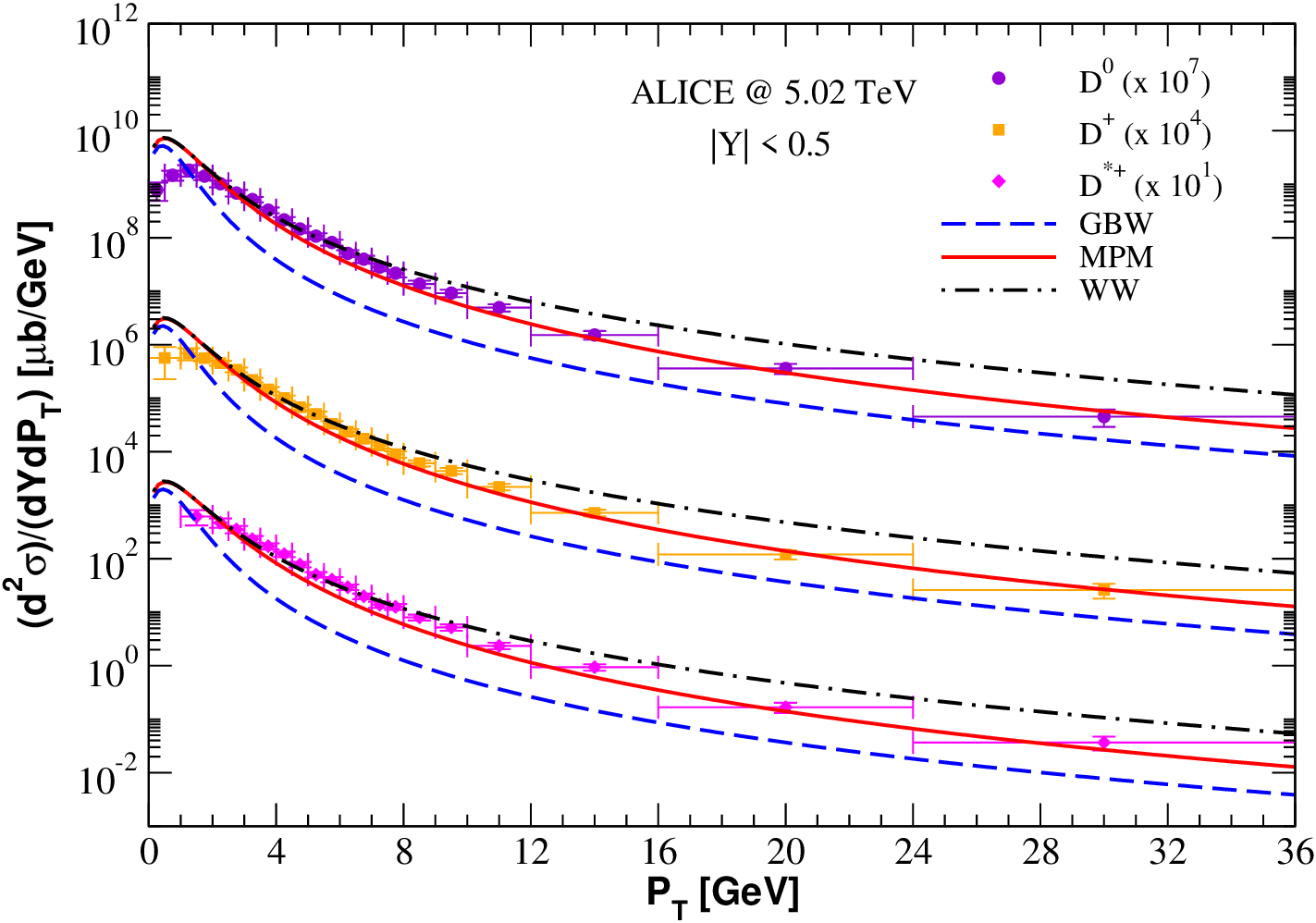}  
\end{tabular}
\caption{Double-differential cross sections for $D^0$, $D^+$, and $D^{*+}$ production in $pp$ collisions at $\sqrt{s} = 5.02$~TeV at midrapidity region. The results are obtained using the GBW, MPM, and WW parametrizations and compared to the experimental measurements from the ALICE experiment \cite{ALICE, ALICE1}.}
\label{pp502}
\end{figure*}

\begin{figure*}[!t]
\begin{tabular}{cc}
\includegraphics[scale=0.35]{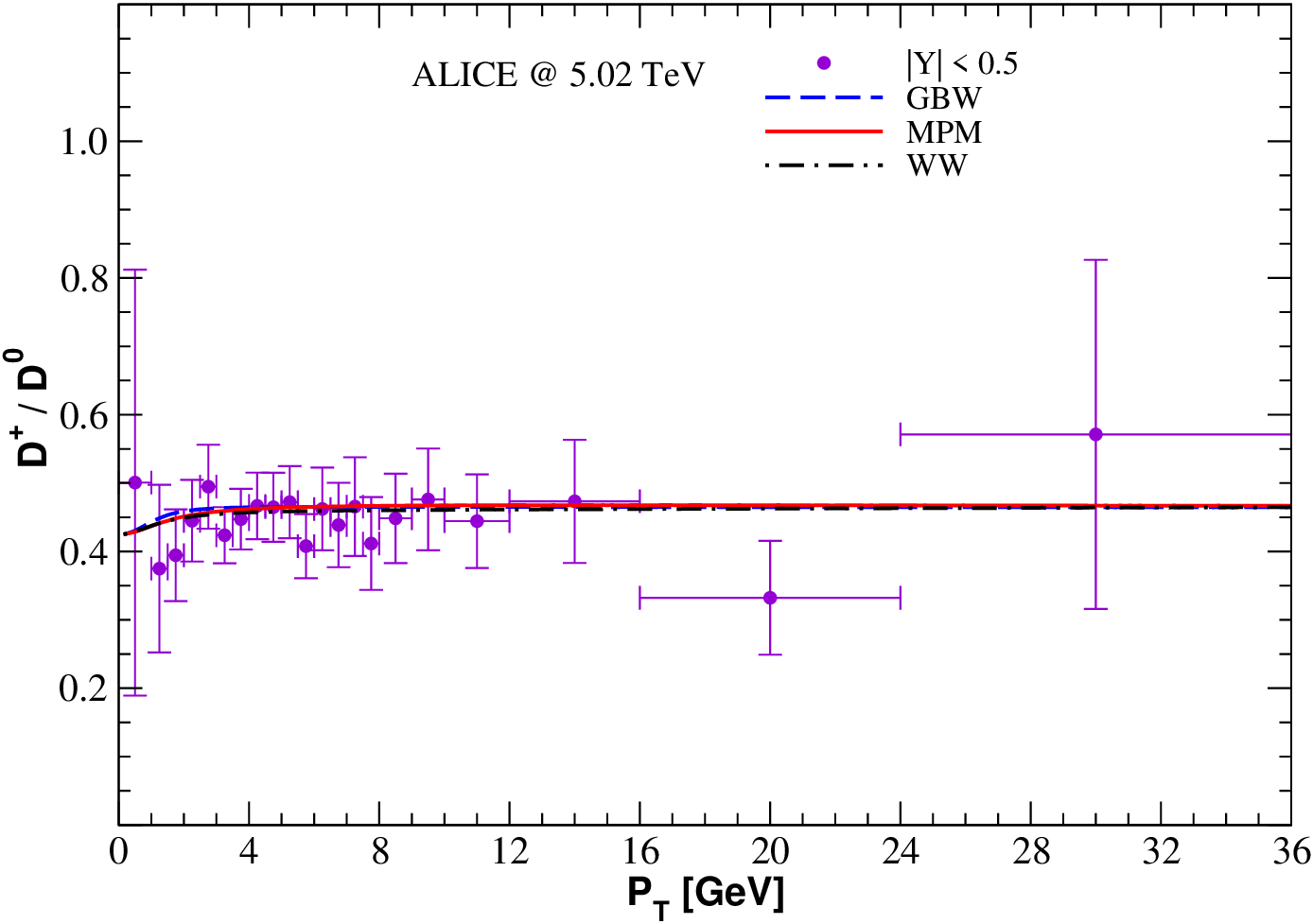}  \includegraphics[scale=0.35]{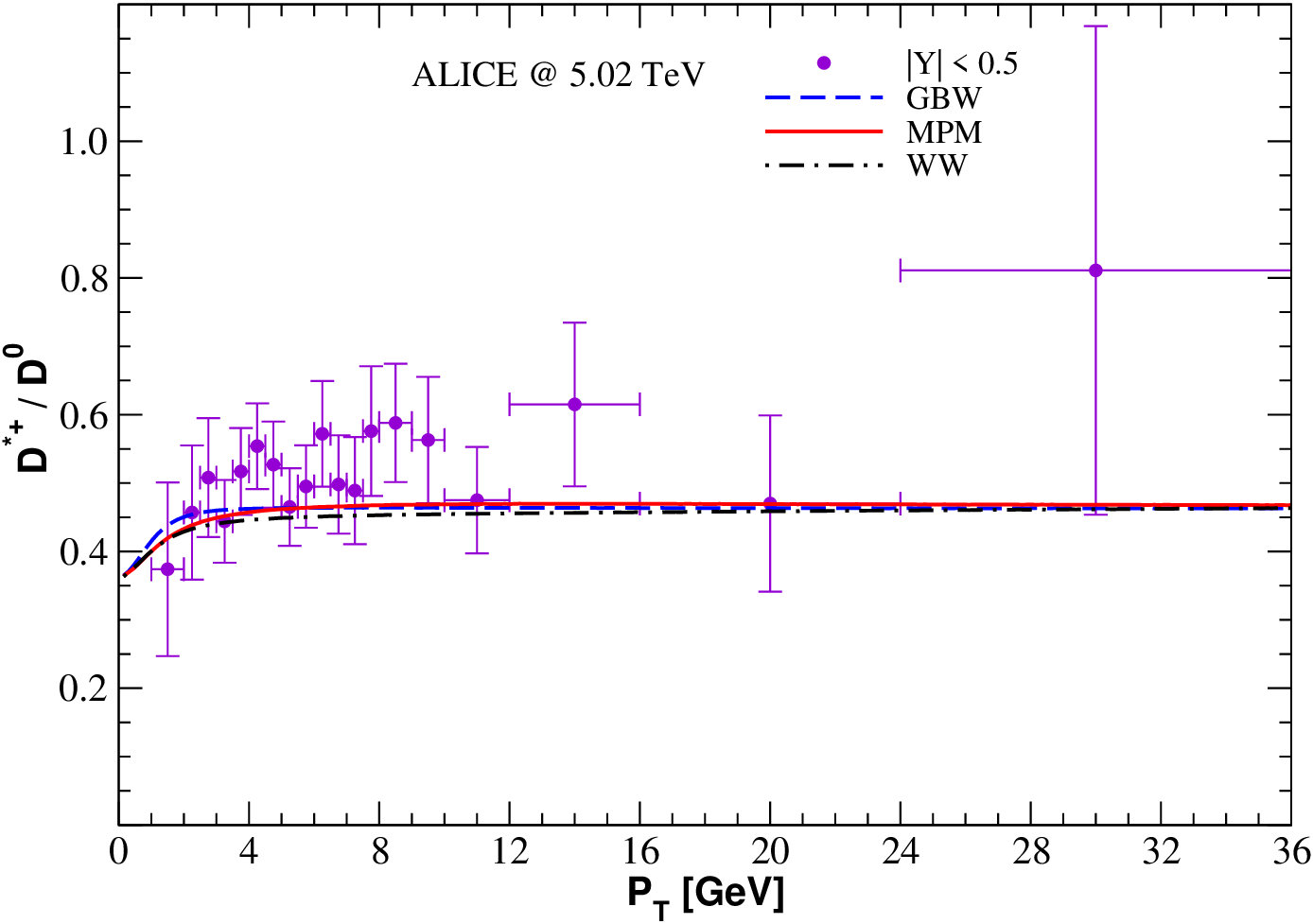} \\
\end{tabular}
\caption{Ratios between the $D^+/D^0$ (left panel) and $D^{*+}/D^0$ (right panel) differential production cross sections in terms of $P_T$. The results are obtained using the GBW, MPM, and WW parametrizations and compared to the experimental measurements from the ALICE experiment \cite{ALICE, ALICE1}}
\label{ratio}
\end{figure*}

In the following an analysis of the $D^0$, $D^+$, and $D^{*+}$ production in $pp$ collisions at midrapidity region is done. The corresponding theoretical predictions for the double-differential cross section at $5.02$~TeV in terms of $P_T$ and $Y$ are displayed in Fig.~\ref{pp502}, where the results are compared with the data collected by the ALICE Collaboration \cite{ALICE, ALICE1}. Apparently, the MPM and WW models reproduce the same results at low $P_T$ spectrum and both predictions are consistent with the experimental data, while the GBW approach gives predictions that underestimate the experimental measurements. We can also observe that the WW estimates begin deviate from the measurements towards of large values of $P_T$ by overshooting them, whereas the MPM results do a better job at describing the data considering the large $P_T$ distribution. Then, we can conclude that the MPM parametrization is able to provide a satisfactory description of the measurements performed by the ALICE experiment at central rapidity. In addition, we present the ratios of the differential cross sections of $D^0$, $D^+$, and $D^{*+}$ mesons produced in $pp$ collisions at $\sqrt{s} = 5.02$~TeV and $Y = 0$ also obtained by the ALICE Collaboration \cite{ALICE, ALICE1}. In particular, the ratios $D^+/D^0$ and $D^{*+}/D^0$ as a function of $P_T$ are shown in Fig.~\ref{ratio}. One can see that the ratios between the corresponding $D$-meson cross section do not provide an evidence of strong $P_T$ dependence, instead showing a constant behavior through the $P_T$ spectrum. This fact indicates that we can not identify discriminatory differences particularly between the fragmentation functions of charm quarks to $D^0$, $D^+$, and $D^{*+}$ mesons. Along with these considerations we can add that the GBW, MPM, and WW predictions are in agreement with the measurements within the experimental uncertainties and we have no basis to distinguish the three UGD parametrizations. In the approximate expression, Eq.~(\ref{approximation}), the ratio $R_{M_1/M_2}$ scales with $\left(\langle z\rangle_{M_1}/\langle z\rangle_{M_2}\right)^{2(1+\lambda)}$ at large $P_T$ and central rapidity $Y=0$.

Besides, the $x_2$ probed in the kinematic ranges determined by the ALICE and LHCb experiments has to be investigated, specially in the very forward ($4 < Y < 4.5$) and central ($|Y| < 0.5$) rapidity bins. We have that the mean value of $\langle x_2 \rangle$ achieved at the LHCb experiment at $5$ and $13$~TeV corresponds to $\langle x_2 \rangle \sim 3\times10^{-5}$ and $\langle x_2 \rangle \sim 2 \times10^{-5}$, respectively, while at the ALICE experiment at $5.02$~TeV this value is $\langle x_2 \rangle \sim 7\times10^{-3}$. Clearly, these results for $\langle x_2 \rangle$ ensure that we perform predictions within the limit of validity of dipole formalism, $x_2\leq 10^{-2}$. 

\begin{figure*}[!t]
\begin{tabular}{cc}
\includegraphics[scale=0.35]{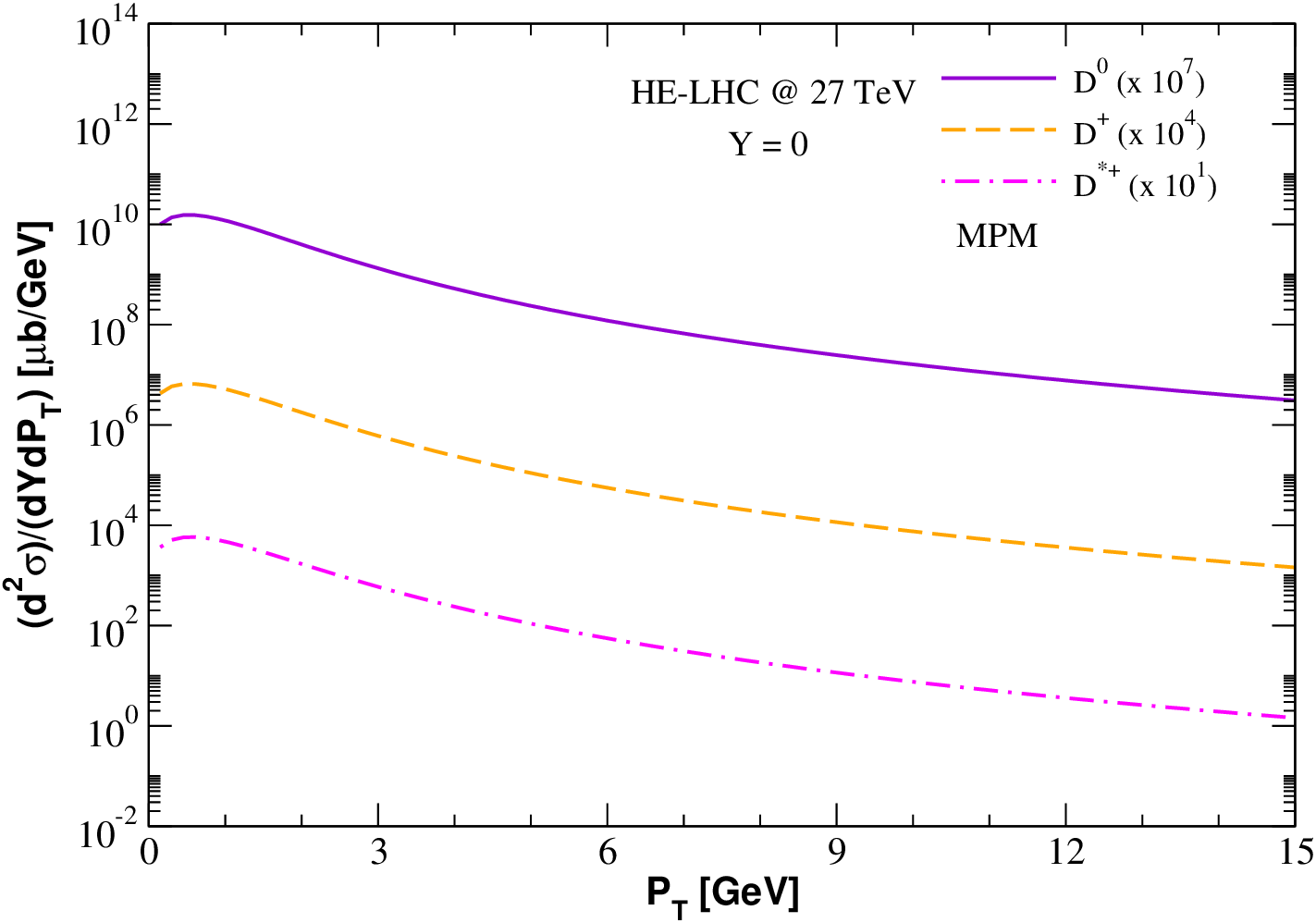}  \includegraphics[scale=0.35]{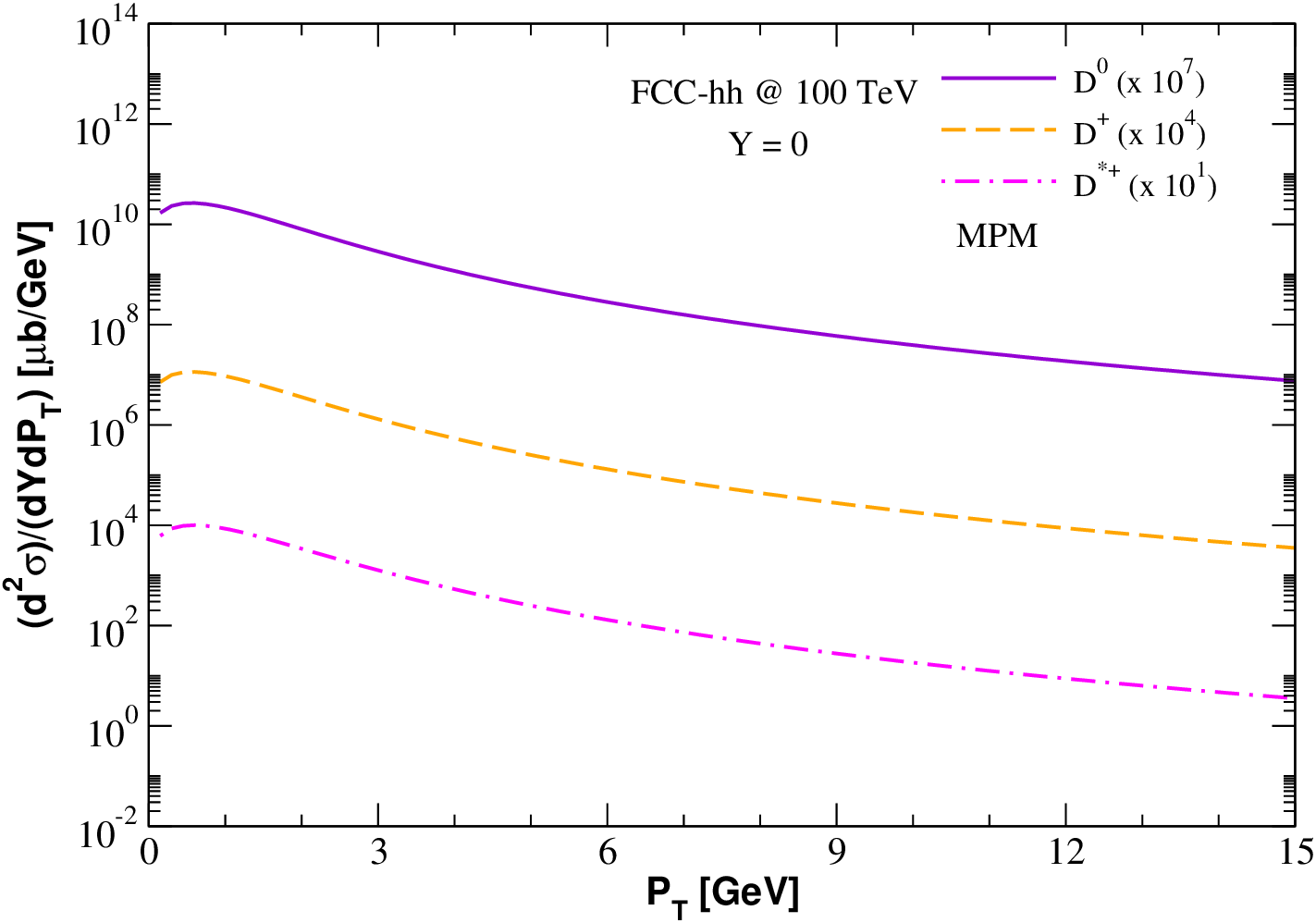} \\
\end{tabular}
\caption{Double-differential cross sections for $D^0$, $D^+$, and $D^{*+}$ production in $pp$ collisions at the HE-LHC ($\sqrt{s} = 27$~TeV, left panel) and FCC-hh ($\sqrt{s} = 100$~TeV, right panel) at midrapidity region. The results are obtained using the MPM parametrization.}
\label{pp27_100}
\end{figure*}

As a matter of completeness, we provide predictions for $D$-meson production concerning the $pp$ collisions aiming the proposals of center-of-mass energies in future colliders. In particular, the High-Energy Large Hadron Collider (HE-LHC) \cite{abada} and the Future Circular Collider (FCC-hh) \cite{abada1} are expected to achieve colliding energies of $\sqrt{s} = 27$~TeV and 100~TeV, respectively. The results corresponding for $D^0$, $D^+$, and $D^{*+}$ with the MPM approach are found in Fig.~\ref{pp27_100}.
Future experimental measurements of $D$-meson production can be fruitful in order to extend the probed kinematic region and to complement our approaches based on QCD dynamics such as color dipole formalism as well as the underlying assumptions considered in the gluon TMD. Along with this aspects the HE-LHC and the FCC-hh could enable us to perform further investigations.

\section{Summary} 
\label{conc}

We investigated the $D$-meson production at high energy $pp$ collisions within the color dipole framework, where we employ three distinct parametrizations for the unintegrated gluon distribution. We provide predictions for the $D^0$, $D^+$, and $D^{*+}$ double-differential cross section that are directly compared to the most recent data reported by the LHC experiments. We have verified that the MPM and WW results are able to satisfactorily describe the transverse momentum and rapidity distributions of the experimental measurements obtained by the ALICE and LHCb Collaborations. On the other hand, we have found out that the GBW parametrization undershoots the experimental data. In particular, better results are given with the MPM approach even at large $P_T$ domain. We have demonstrated that the treatment of the $D$-meson production at high energies can be appropriately formulated in the color dipole framework where the corresponding results are parameter free.

In view of the trend found in the MPM and WW results obtained in this analysis, the new data taking from the future colliders in $pp$ mode will be valuable to extend the kinematic region and to improve the MPM and WW parametrizations for the unintegrated gluon distribution.

\section*{Acknowledgements}

This work was partially financed by the Brazilian funding agencies CAPES, CNPq, and FAPERGS. This study was financed in part by the Coordena\c{c}\~ao de Aperfei\c{c}oamento de Pessoal de N\'{\i}vel Superior - Brasil (CAPES) - Finance Code 001. GGS
acknowledges funding from the Brazilian agency Conselho Nacional
de Desenvolvimento Científico e Tecnológico (CNPq) with grant
311851/2020-7.

\end{document}